\documentclass[12pt]{article}
\usepackage{epstopdf}
\usepackage{subfigure}
\usepackage{mathrsfs}
\usepackage{amssymb}
\usepackage{amsmath}
 \usepackage{amsfonts}
\usepackage[T1]{fontenc}
\usepackage{multirow}
\usepackage{pxfonts}
\usepackage{epsfig}
\usepackage{color}
\usepackage{ulem}

\newtheorem{theorem}{Theorem}[section]

\newtheorem{proposition}[theorem]{Proposition}

\begin{document}

\renewcommand{\baselinestretch}{1.5}

\fontsize{10.95}{14pt plus.8pt minus .6pt}\selectfont
\vspace{0.8pc}
\centerline{\bf On finite sample properties of nonparametric discrete asymmetric kernel estimators}
\vspace{2pt}
\centerline{To appear in {\it Statistics: A Journal of Theoretical and Applied Statistics}, 2017}
\vspace{.4cm}
\centerline{Tristan Senga Kiess\'e}
\vspace{.2cm}
\centerline{UMR SAS, INRA, Agrocampus Ouest, F-35000 Rennes, France.}
\vspace{.1cm}
\centerline{\it tristan.senga-kiesse@inra.fr}
\vskip .55cm

\begin{center}
 {\bf ABSTRACT}
\end{center}
The discrete kernel method was developed to estimate count data distributions, distinguishing discrete associated kernels based on their asymptotic behaviour.  This study investigates the class of discrete asymmetric kernels and their resulting non-consistent estimators, but this theoretical drawback of the estimators is balanced by some interesting features in small/medium samples. The role of modal probability and variance of discrete asymmetric kernels is highlighted to help better understand the performance of these kernels, in particular how the binomial kernel outperforms other asymmetric kernels. The performance of discrete asymmetric kernel estimators of probability mass functions is illustrated using simulations, in addition to applications to real data sets.

\vspace{9pt}

\section{Introduction}

 The concept of {\it discrete associated kernels} was introduced to define discrete non/semi-parametric kernel estimators of probability mass functions (p.m.f.) or count regression functions on a discrete support $\mathcal{S}$ as a non-negative integer set $\mathbb{N}$ \cite{KokonSenga11,Senga08}. For instance, the discrete kernel estimator $\widetilde{f}$ of an unknown p.m.f. $f$ of i.i.d. observations $(X_{i})_{i=1,\cdots,n}$ was constructed to behave asymptotically as the frequency estimator $\widetilde{F}(x)=n^{-1}\sum_{i=1}^{n}{\bf 1}_{\{x\}}(X_{i}),  x\in\mathcal{S}$, where ${\bf1}_{A}$ denotes the indicator function of the set A (for details about $\widetilde{f}$, see later equation (\ref{eq:estim_fn}) in Section 3). Indeed, the estimator $\widetilde{F}$ had long been regarded as the nonparametric reference for count data with large sample sizes. Then, the discrete kernel estimator $\widetilde{f}$ was introduced to provide an alternative to $\widetilde{F}$ for modelling the p.m.f. $f$ of count data \cite{KokonSenga11}. To this end, the estimator  $\widetilde{f}$ has a bandwidth parameter $h>0$ which serves to control the quality of adjustment of the p.m.f. $f$ estimate, in contrast to the frequency estimator $\widetilde{F}$ of $f$ using Dirac type kernel $D_{x}={\bf 1}_{\{x\}}$, for which $h=0$. Thus, one uses the terms {\it smoothness} or {\it smoothing} even though one talks about a discrete p.m.f. In summary,  the discrete associated kernel approach extends the continuous kernel estimation procedure \cite{Simonoff96,Tsyb04} to the modelling of count data distributions. Aitchison-Aitken \cite{AitchAitk76} may be cited among the seminal works on discrete kernels. Studies using the discrete associated kernel method are now focused on the Bayesian approach for bandwidth choice, e.g. \cite{Senga16,Zougab13}, or the multivariate case, e.g. \cite{Belaid16}. 

\begin{figure}[!h]
\begin{center}
\includegraphics[width=\textwidth]{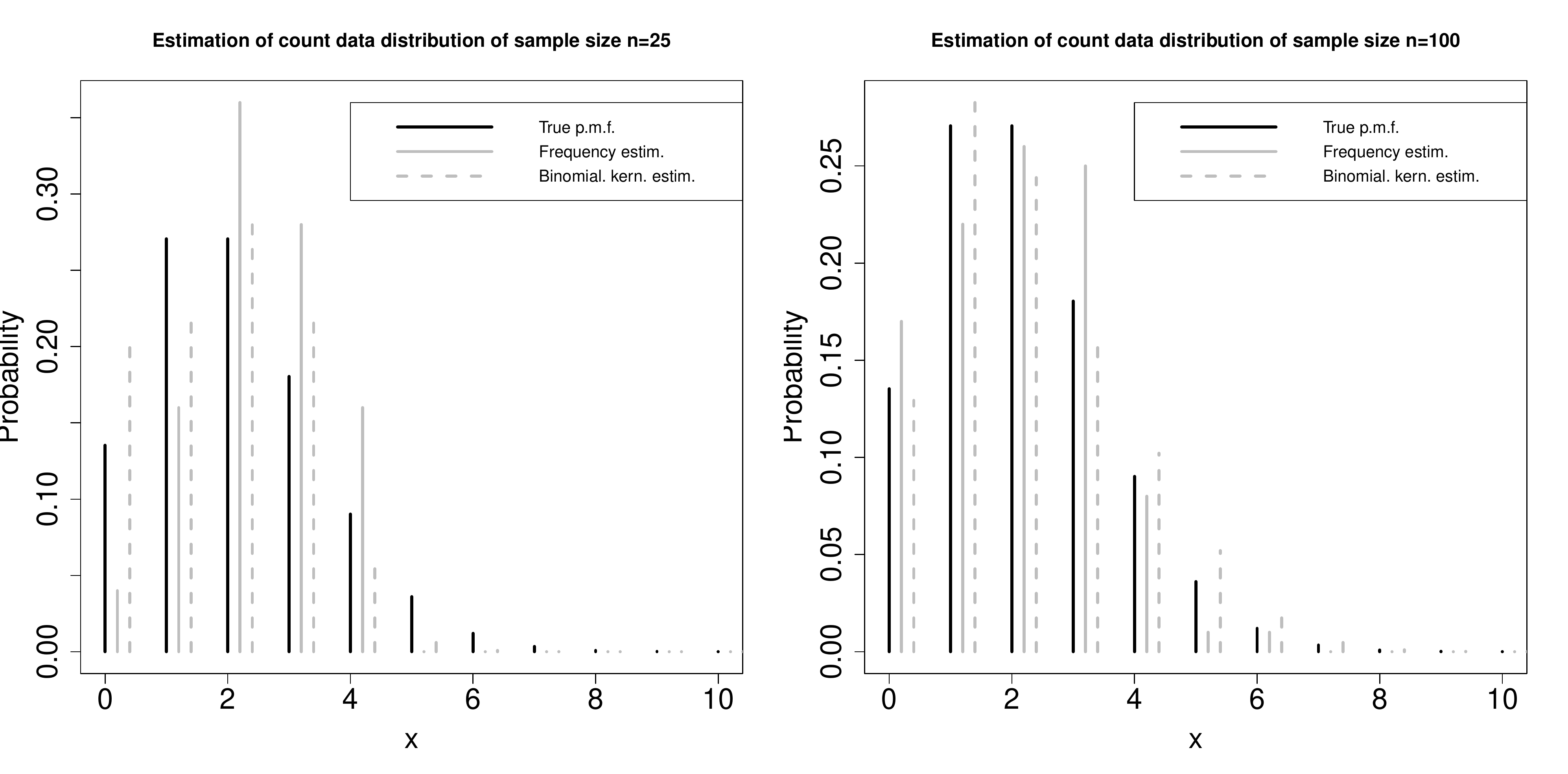}
\end{center}
\caption{{\color{black}{Exemplary run of}} estimation of count data  from a Poisson distribution of mean $\mu=5$ by nonparametric kernel and frequency estimators }
\label{fig:estimation}
\end{figure}

Two classes of discrete associated kernels were proposed depending on whether they tend asymptotically to the Dirac type kernel or not.
One class of kernels contains discrete triangular kernels \cite{KokonZocchi10} and  Aitchison-Aitken \cite{AitchAitk76} and Wang-van Ryzin \cite{WangVan81} kernels (examples 3 and 4 in \cite{KokonSenga11}), which tend asymptotically to the Dirac type kernel. The nonparametric estimator of a p.m.f. using this type of discrete kernels is consistent. The other class of kernels contains discrete standard asymmetric kernels constructed from usually discrete probability distributions such as Poisson, binomial and negative binomial. The nonparametric estimator of a p.m.f. using discrete standard kernels does not tend asymptotically to the frequency estimator, but it was shown to be useful for estimating small/medium sample sizes. \cite{KokonSenga11} For example, an estimator using a standard (binomial) kernel outperforms the frequency estimator for count data, {\color{black}{when simulating $250$ replicates}} of sample sizes $n=\{25, 100\}$ from a Poisson distribution (Figure \ref{fig:estimation} and Table \ref{tab:estimation}). Thus, it is worth studying non-consistent discrete standard kernel estimators in a situation like this, in which other consistent estimators abound. 

\begin{table}[ht]
\caption{{\color{black}{Average}} integrated squared errors for frequency estimator $\widetilde{F}$ and discrete kernel estimator $\widetilde{f}$ of a simulated count data distribution $f$}
\scriptsize
\center
{\begin{tabular}[l]{@{}ccc}\hline
Sample size $n$& $(1/250)\sum_{j=1}^{250}[\sum_{x\in\mathbb{N}}\{f(x)-\widetilde{f}(x)\}^{2}]$  &  $(1/250)\sum_{j=1}^{250}[\sum_{x\in\mathbb{N}}\{f(x)-\widetilde{F}(x)\}^{2}]$\\ \hline
$25$& $0.0099$ &$0.0320$  \\
$100$& $0.0023$& $0.0086$ \\ \hline
\end{tabular} }
\label{tab:estimation}
\end{table}

The present work supplements the existing literature on discrete associated kernel estimation \cite{KokonSenga11,Senga08}. In particular, the study aims to (i) help understand the finite-sample performance of discrete standard kernels and (ii) highlight the utility of non-consistent discrete standard kernel estimators.  To this end, the modal probability and variance of discrete standard kernels are presented in a common form useful for comparing their relative efficiencies. Compared to existing studies, this study examines how the binomial kernel outperforms other asymmetric kernels  (section \ref{sec2}). Then, an approximate global squared error of the discrete kernel estimator is derived, and the performance of nonparametric estimators using discrete standard asymmetric kernels is ranked according to the error criterion considered (section \ref{sec3}). Finally, the performance of non-consistent discrete standard kernel estimators is illustrated for simulated and real count data sets and compared to a consistent discrete associated kernel estimator and/or the frequency estimator  (section \ref{sec4}).

\section{Discrete kernels}\label{sec2}

This section presents the two classes of kernels mentioned previously. The first subsection recalls the expressions which characterize a discrete associated kernel. The second subsection proposes new expressions to characterize discrete standard asymmetric kernels for deeper investigation of their properties. Hereafter, the support $\mathcal{S}$ of the p.m.f. to estimate is assumed to be the non-negative integer set $\mathbb{N}$.

\subsection{Discrete associated kernel}

Let us consider a fixed point $x\in\mathbb{N}$ and a bandwidth parameter $h>0$. The discrete kernel $K_{x,h}$ is associated with a r.v. $\mathcal{K}_{x,h}$, i.e. $K_{x,h}(y)=\Pr(\mathcal{K}_{x,h}=y)$, on support  $\mathcal{S}_{x}$ which contains $x$. The main property of $K_{x,h}$ can be summarised in the following behaviour of its modal probability:
\begin{equation}\label{eq:ModalProb}
\Pr(\mathcal{K}_{x,h}=x) \to \Pr(\mathcal{D}_{x}=x)=1 \text{ \ as \ } h\to 0,
\end{equation}
with $\mathcal{D}_{x}$ being a r.v. of p.m.f  the Dirac type kernel $D_{x}$ on support $\mathcal{S}_{x}=\{x\}$. The idea is that the discrete associated kernel must attribute the more important probability mass (i.e. closest to one) at target $x\in\mathbb{N}$, while having a smoothing parameter $h>0$ to take into account the probability mass at points $y\in\mathbb{N}\setminus\{x\}$ in the neighboorhood of $x$. The following {\color{black}{expressions}} of $\mathcal{K}_{x,h}$'s expectation and variance result from equation (\ref{eq:ModalProb}):
\begin{equation*}
(\text{E}_{1}): \textnormal{E}(\mathcal{K}_{x,h})=x + a(x,h) \text{ \ and \ } (\text{E}_{2}): \textnormal{Var}(\mathcal{K}_{x,h})=b(x,h),
\end{equation*}
where both $a(x,h)$ and $b(x,h)$ tend to $0$ as $h$ goes to $0$, since $K_{x,h}(x) \to 1$ and, for $y\neq x$, $K_{x,h}(y) \to 0$ as $h$ goes to $0$.\cite{KokonSenga11} 

We now describe how the previous expressions were obtained, details not completely presented in most existing references.  The expressions $(\text{E}_{1})$ and $(\text{E}_{2})$ resulted from developing the kernel's expectation and variance around target $x$ as:
\begin{equation*}
\textnormal{E}(\mathcal{K}_{x,h})=xK_{x,h}(x) + \sum_{y\neq x}yK_{x,h}(y) = x + x\{K_{x,h}(x) - 1\} + \sum_{y\neq x}yK_{x,h}(y) 
\end{equation*}
and 
\begin{eqnarray*}
\textnormal{Var}(\mathcal{K}_{x,h})&=&  \sum_{y\in \mathcal{S}_{x}}y^{2}K_{x,h}(y) - \big\{\sum_{y\in \mathcal{S}_{x}}yK_{x,h}(y)\big\}^{2}\\
&=& x^{2}K_{x,h}(x) -  x^{2}K^{2}_{x,h}(x) + \sum_{y\neq x}y^{2}K_{x,h}(y) + x^{2}K_{x,h}(x) - \big\{\sum_{y\in \mathcal{S}_{x}}yK_{x,h}(y)\big\}^{2}\\
&=& x^{2}K_{x,h}(x)\{1- K_{x,h}(x)\} + q(x,h),
\end{eqnarray*}
with $$  q(x,h)= \sum_{y\neq x}y^{2}K_{x,h}(y) + x^{2}K_{x,h}(x) - \big\{\sum_{y\in \mathcal{S}_{x}}yK_{x,h}(y)\big\}^{2}\to 0 \text{\ when\ } h\to 0.$$

For $x \in \mathbb{N}$ and $h>0$, an example of discrete associated kernel is the symmetric triangular kernel $K_{p;x,h}$ associated with the r.v. $\mathcal{K}_{p;x,h}$ as $K_{p;x,h}=\Pr(\mathcal{K}_{p;x,h}=y)$, for $y\in\mathcal{S}_{p;x}=\{x,x\pm1,...,x\pm p\}$. The p.m.f. of  $K_{p;x,h}$ is given by
\begin{equation*}\label{TriangKern}
K_{p;x,h}(y) = \frac{(p+1)^{h}-|y-x|^{h}}{(2p+1)(p+1)^{h}-2\sum_{k=0}^{p}k^{h}},\text{ \ } p\in\mathbb{N}.
\end{equation*}
Its modal probability and variance can be developed as follows:
\begin{equation*}
(\text{A}_{1}):\Pr(\mathcal{K}_{p;x,h}=x)=1-2h\text{A}(p)+ O(h^2) \text{ \ and \ } 
(\text{A}_{2}):\text{Var}(\mathcal{K}_{p;x,h})=2h\text{V}(p)+ O(h^{2}),
\end{equation*}
with $\text{A}(p)=p\log(p+1)-\sum_{k=1}^{p}\log(k)$ and $\text{V}(p)=\{p(2p^{2}+3p+1)/6\}\log (p+1)-\sum_{k=1}^{p}k^{2}\log(k)$. \cite{Senga14} Thus, the expression of modal probability in equation ($\text{A}_{1}$) quickly shows that equation (\ref{eq:ModalProb}) is verified by this discrete associated kernel.  The expansions ($\text{A}_{1}$)-($\text{A}_{2}$) of  modal probability and variance of $K_{p;x,h}$ will be useful for comparison with discrete standard asymmetric kernels (next section).

\subsection{Discrete standard kernels}
This subsection focuses on the discrete asymmetric kernels constructed from binomial, Poisson and negative binomial distributions  \cite{KokonSenga11,Senga08} and which do not satisfy equation (\ref{eq:ModalProb}). In particular, we provide new expressions of the modal probability and variance of the discrete asymmetric kernels when considering $h\to0$, which allows the modal probability and variance of these kernels to be compared. 

\subsubsection{Poisson kernel} For $x\in\mathbb{N}$ and $h>0$, the Poisson kernel $P(x;h)$ derived from the Poisson distribution $\mathcal{P}(x+h)$ associated with the r.v. $\mathcal{P}_{x,h}$ as $P(x;h)(y)=\Pr(\mathcal{P}_{x,h}=y)$, for $y\in\mathcal{S}_{x}=\mathbb{N}$.  The modal probability  of Poisson kernel  using a Taylor expansion of second order at $h\to 0$ can be obtained as  
\begin{eqnarray*}\label{eq:modprob_pois}
\Pr(\mathcal{P}_{x,h}=x)=\frac{x^{x}\exp(-x)}{x!}\biggl( 1+\frac{h}{x}\biggr)^{x}\exp(-h) &=&(1-h^{2})\frac{x^{x}\exp(-x)}{x!} + O(h^{2}) \\
&=&(1-h^{2})P(x;0)(x) + O(h^{2})
\end{eqnarray*}
 and its variance is given by $\text{Var}(\mathcal{P}_{x,h})=x+h${\color{black}{, with $P(x;0)(x)=x^{x}\exp(-x)/x!$ being the modal probability at target $x$ when $h\to 0$.}}

\subsubsection{Binomial kernel.}  For $x\in\mathbb{N}$ and $h\in(0,1]$, the binomial kernel $B(x;h)$
is constructed from the binomial distribution $\mathcal{B}\{x+1,(x+h)/(x+1)\}$ associated with the r.v. $\mathcal{B}_{x,h}$ on $\mathcal{S}_{x}=\{0,1,\cdots,x+1\}$ such that
\begin{eqnarray*} \label{eq:modprob_bin}
\Pr(\mathcal{B}_{x,h}=x)=(1-h)x^{x}\biggl(\frac{1+h/x}{x+1}\biggr)^{x}&=&(1-h^{2})\biggl(\frac{x}{x+1}\biggr)^{x} + O(h^{2})\\
&=& (1-h^{2})B(x;0)(x) + O(h^{2})
\end{eqnarray*}
and $\text{Var}(\mathcal{B}_{x,h})=x/(x+1) + h\bigl\{(1-x)/(x+1)\bigr\} - h^{2}/(x+1)${\color{black}{, with $B(x;0)(x)$ being the modal probability at target $x$ when $h\to 0$}}.

\subsubsection{Negative binomial kernel}  For $x\in\mathbb{N}$ and $h>0$, the negative binomial $NB(x;h)$ derived from
the negative binomial distribution $\mathcal{NB}\{x+1,(x+1)/(2x+1+h)\}$  associated with the r.v. $\mathcal{NB}_{x,h}$ on $\mathcal{S}_{x}=\mathbb{N}$. {\color{black}{Its modal probability can be expressed as}} 
\begin{eqnarray*}
\Pr(\mathcal{NB}_{x,h}=x)&=&\frac{(2x)!}{(x!)^{2}}\biggl(\frac{x}{2x+1}\biggr)^{x}\biggl(\frac{x+1}{2x+1}\biggr)^{x+1}\frac{(1+h/x)^{x}}{\{1+h/(2x+1)\}^{2x+1}}\notag \\
&=&(1-h^{2})\frac{(2x)!}{(x!)^{2}}\biggl(\frac{x}{2x+1}\biggr)^{x}\biggl(\frac{x+1}{2x+1}\biggr)^{x+1} + O(h^{2})\\ \label{eq:modprob_binneg}
&=&(1-h^{2})NB(x;0)(x) + O(h^{2})
\end{eqnarray*}
 and  $\text{Var}(\mathcal{NB}_{x,h})=x+ x^2/(x+1) + h\bigl\{2x/(x+1) +1 \bigr\} + h^{2}/(x+1)${\color{black}{, with $NB(x;0)(x)$ being the modal probability at target $x$ when $h\to 0$.}}

\medskip
We propose a generalization of the behaviour of these standard kernels through the following assumptions on both their probability at target $x$ and variance:
$$(\text{A}_{3}):\Pr(\mathcal{K}_{x,h}=x)=(1-h^{2})K_{x,0}(x) + O(h^{2})$$ and  
$$(\text{A}_{4}):\text{Var}(\mathcal{K}_{x,h})=V_{\mathcal{K}_{x,h}}(x) + hU_{\mathcal{K}_{x,h}}(x) + O(h^{2}),$$
where $$\sum_{y\in\mathcal{S}_{x}\setminus\{x\}}\Pr(\mathcal{K}_{x,h}=y)=1-(1-h^2)K_{x,0}(x) + O(h^{2}),  $$ with the terms $V_{\mathcal{K}_{x,h}}$ and $U_{\mathcal{K}_{x,h}}$ depending on the discrete kernel used. As $h \to 0$,  the modal probability and variance of discrete standard asymmetric kernels are such that $\Pr(\mathcal{K}_{x,h}=x)\rightarrow K_{x,0}(x)\neq1$ and $\text{Var}(\mathcal{K}_{x,h})\to V_{\mathcal{K}_{x,h}}(x)\neq 0, \text{ } x\in\mathbb{N}\setminus\{0\}$. Unlike the assumptions $(\text{A}_{1})$-$(\text{A}_{2})$ for discrete symmetric triangular kernels, the assumptions $(\text{A}_{3})$-$(\text{A}_{4})$ do not satisfy equation (\ref{eq:ModalProb}) for discrete standard kernels, which explains the main difference between these two classes of kernels.\\

\textit{Remark 1.} \textit{(i)} The discrete standard asymmetric kernels were originally constructed such that their expectation and variation must satisfy 
\begin{equation*}
\textnormal{E}(\mathcal{K}_{x,h})=x + h \text{ \ and \ } \lim_{h\to 0}\textnormal{Var}(\mathcal{K}_{x,h}) \in \mathcal{V}(0),
\end{equation*}
with $\mathcal{V}(0)$ a set in the neighborhood of $0$, different from discrete associated symmetric kernels, for which $\textnormal{E}(\mathcal{K}_{x,h})=x$.

\begin{figure}[!h]
\begin{center}
\includegraphics[height=200pt,width=410pt]{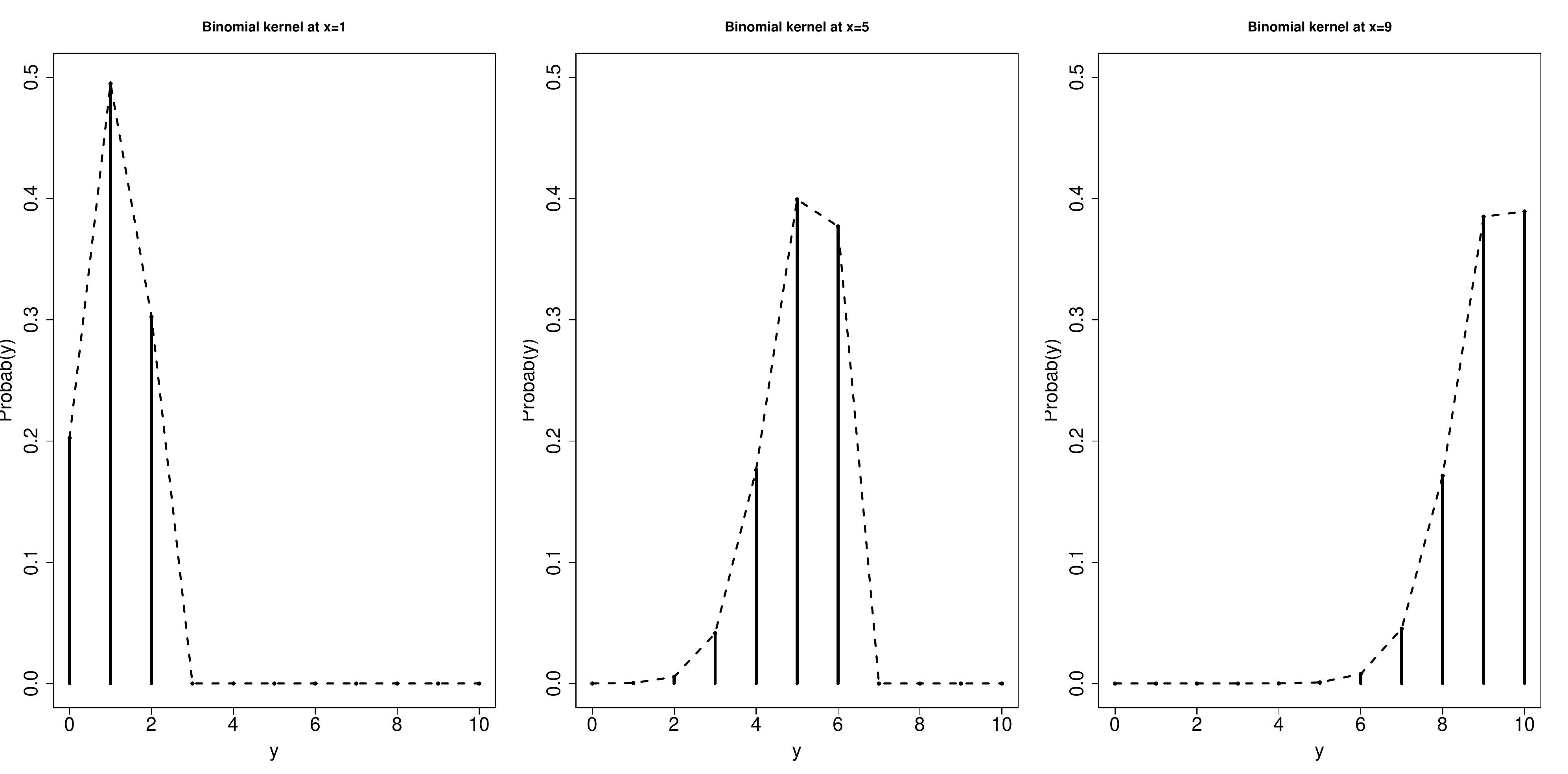}
\end{center}
\caption{Shape of binomial kernel at various targets $x$ for fixed $h=0.1$ on support $\mathcal{S}_{x}=\{0,1,\ldots,10\}$.}
\label{fig:shape_binom_kern}
\end{figure}

\textit{(ii)} The discrete standard  asymmetric kernels take advantage of their variable asymmetric shape (e.g., Figure \ref{fig:shape_binom_kern}), similar to that of asymmetric continuous kernels \cite{Chen99,HagmScail07}. This shape is adaptive depending on the estimation target $x$, which makes these kernels useful for the boundary bias problem.
 
\subsubsection{Comparison of discrete standard kernels}  \label{ssec2:comparaison_kernel}                                            
Under the common assumptions $(\text{A}_{3})$-$(\text{A}_{4})$, we compare the discrete standard kernels on the basis of their modal probability and variance. 

\begin{figure}[!h]
\begin{center}
\includegraphics[width=\textwidth]{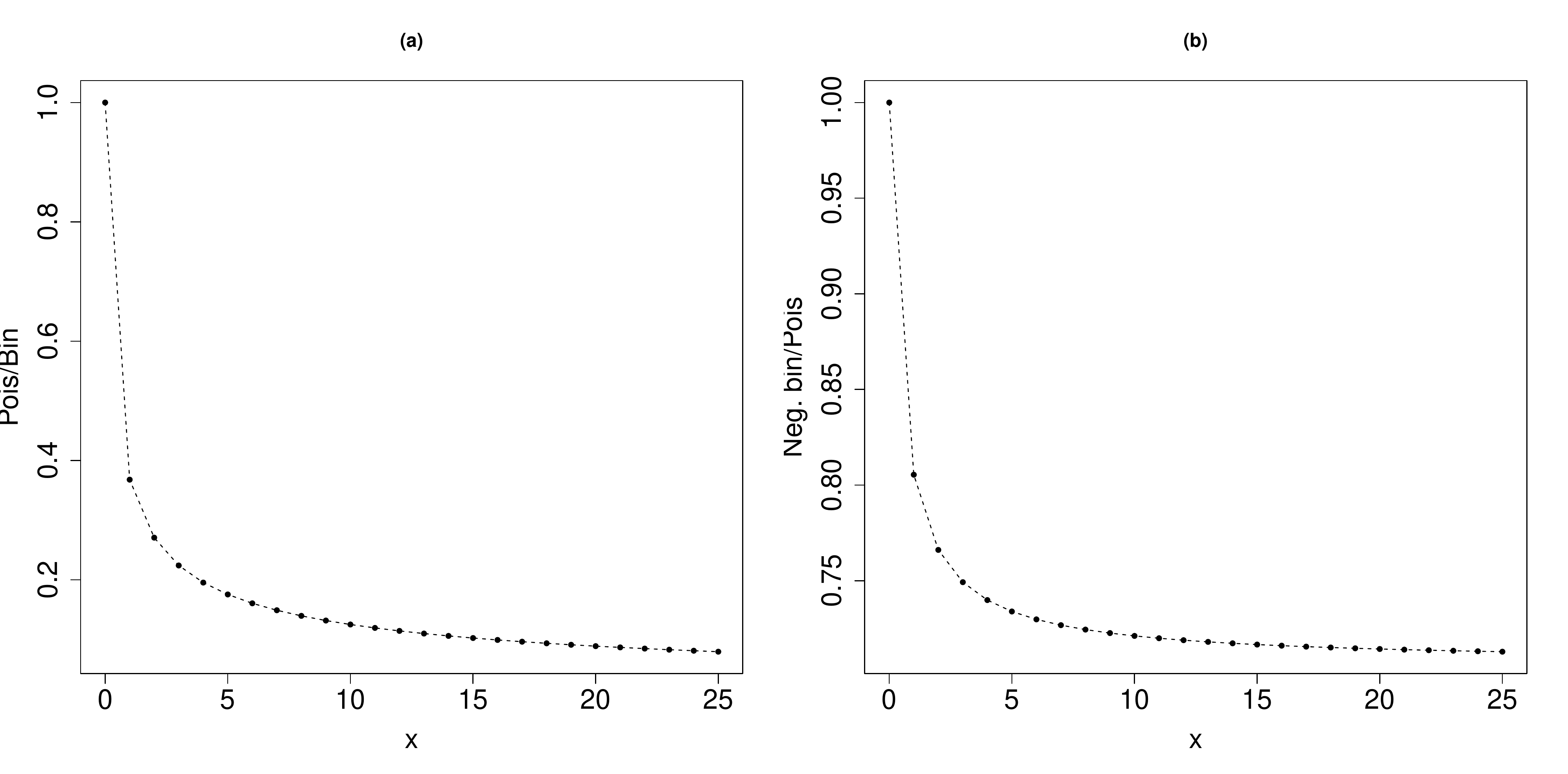}
\end{center}
\caption{Graph of ratios of main terms in modal probability of Poisson by binomial kernels (a) and negative binomial by Poisson kernels (b) }
\label{fig:quotient_prob}
\end{figure}

\begin{figure}[!h]
\begin{center}
\includegraphics[height=400pt,width=400pt]{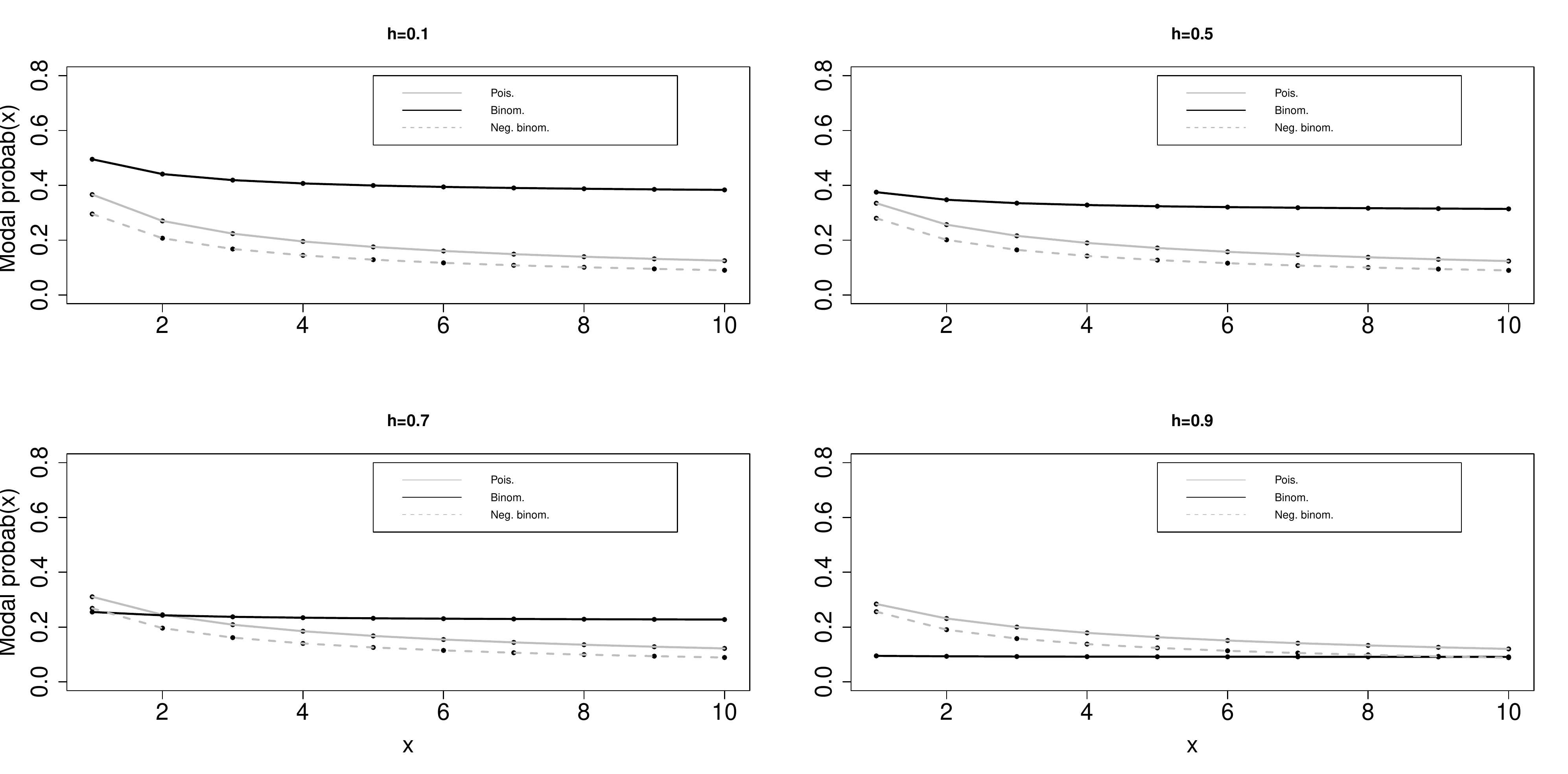}
\end{center}
\caption{Graph of modal probability of discrete standard kernels for some values of $h\in(0,1]$ and $x=1,2,\ldots,10$ }
\label{fig:modal prob}
\end{figure}

For $x\in\mathbb{N}$, we first focus on the modal probability of discrete standard kernels through the terms $K_{x,0}(x)$ in the expression $(\text{A}_{3})$. For Poisson and binomial kernels, we obtain
\begin{equation}\label{eq:r1}
r_{1}(x)=\frac{P(x;0)}{B(x;0)}=\frac{(x+1)^{x}\exp(-x)}{x!}\leq1;
\end{equation}
and, for Poisson and negative binomial kernels, we obtain
\begin{equation}\label{eq:r2}
r_{2}(x)=\frac{NB(x;0)}{P(x;0)}=\frac{(2x)!}{(x!)^{2}}\biggl(\frac{x}{2x+1}\biggr)^{x}\biggl(\frac{x+1}{2x+1}\biggr)^{x+1}\times\frac{x!\exp(x)}{x^{x}}\leq1.
\end{equation}
Figure \ref{fig:quotient_prob} plots the ratio functions $r_{1}(x)$ and $r_{2}(x)$. As $h \to 0$, the following ranking occurs for the main terms in modal probability of discrete standard kernels: $NB(x;0)\leq P(x;0) \leq B(x;0)$, $x\in\mathbb{N}$. However, this ranking is not always available for all $h$-values. For instance, for chosen $h$-values in $(0,1]$ and $x=2,\ldots,10$, the modal probability of the binomial kernel is larger than those of Poisson and negative binomial kernels, except for $h=0.9$ (Figure \ref{fig:modal prob}).  Thus, a maximum bandwidth $h_{0}>0$ exists such that, for $h<h_{0}$, the binomial kernel attributes the largest probability mass at target $x\in\mathbb{N}$, unlike the two other discrete standard kernels. In contrast, for $h>h_{0}$, the Poisson and negative binomial kernels can attribute more probability mass at $x\in\mathbb{N}$ than the binomial kernel.  Conversely, the previous remark implies that a maximum sample size $n_{0}$ exists such that for $n<n_{0}$ the Poisson and negative binomial kernels can attribute more probability mass at $x\in\mathbb{N}$ than the binomial kernel (and reciprocally), since the smoothing parameter $h=h(n)$ is linked to the sample size $n$ such that $h\to0$ when $n \to \infty$. The main question thus remains to find the maximum $h_{0}$-value (or reciprocally the maximum $n_{0}$-sample size). These observations will be illustrated later using simulations (section \ref{sec4}).

Ultimately, we formulate the following proposition on the basis of the above.

\begin{proposition}\label{Propostion1}
Consider any fixed $x\in\mathbb{N}$ and $h>0$. Under assumptions $(\text{A}_{3})$-$(\text{A}_{4})$,  as $h\to 0$ , the modal probability and variance of the three discrete standard asymmetric kernels satisfy:
\begin{equation}\label{eq:probaMod}
\Pr(\mathcal{NB}_{x,h}=x)\leq\Pr(\mathcal{P}_{x,h}=x)\leq\Pr(\mathcal{B}_{x,h}=x)
\end{equation}
and
\begin{equation}\label{eq:varianceKern}
\textnormal{Var}(\mathcal{NB}_{x,h})\geq\textnormal{Var}(\mathcal{P}_{x,h})\geq\textnormal{Var}(\mathcal{B}_{x,h}).
\end{equation}
\end{proposition}

{\bf Proof}. The comparison of the modal probability of kernels in equation (\ref{eq:probaMod}) comes from equations (\ref{eq:r1}) and (\ref{eq:r2}).  

For equation (\ref{eq:r1}), we show that the ratio $r_{1}$ is decreasing with respect to $x\in\mathbb{N}$ and less than $1$. To this end, by using a Taylor expansion as $x\to\infty$, we successively express:
\begin{eqnarray*}
\ln\left\{ \frac{r_{1}(x+1)}{r_{1}(x)} \right\}&=&\ln\left\{\frac{(x+2)}{(x+1)}^{x+1}\exp(-1)\right\}\\
&=&(x+1)\ln\left(1+\frac{1}{x+1}\right) - 1\\
&\approx&(x+1)\left(\frac{1}{x+1} - \frac{1}{2(x+1)^2} \right) - 1<0.\\
\end{eqnarray*} 
Hence, we obtain $r_{1}(x+1)\leq r_{1}(x)$ with $r_{1}(0)=1$.

Now, we focus on the ratio $r_{2}$ in  equation (\ref{eq:r2}). Without providing all calculation details, we first obtain
\begin{eqnarray*}
\frac{r_{2}(x+1)}{r_{2}(x)}&=&2\frac{(2x+1)^{2x+1}}{\{2(x+1)+1\}^{2(x+1)+1}}\times\left(\frac{x+2}{x+1}\right)^{x+1}\times(x+2)\times\exp(1)\\
&=&2\frac{(2x+1)^{2x+1}}{\{(2x+1)+2\}^{2x+1}}\times\left(1+\frac{1}{x+1}\right)^{x+1}\times\frac{(x+1)+1}{2(x+1)+1}\times\frac{\exp(1)}{2x+3}\\
&=&\frac{1}{\{1+2/(2x+1)\}^{2x+1}}\times\left(1+\frac{1}{x+1}\right)^{x+1}\times\frac{1+1/(x+1)}{1+1/2(x+1)}\times\frac{\exp(1)}{2x+3}.\\
\end{eqnarray*} 
Then, by using a Taylor expansion as $x\to\infty$, we express
\begin{eqnarray*}
\ln\left\{\frac{r_{2}(x+1)}{r_{2}(x)}\right\}&=& - (2x+1)\ln\left(1+\frac{2}{2x+1}\right)+(x+1)\ln\left(1+\frac{1}{x+1}\right)\\
&& +\ln\left(1+\frac{1}{x+1}\right)-\ln\left(1+\frac{1}{2(x+1)}\right) - \ln(2x+3) + 1\\
&\approx&- 2 + 1 + \frac{1}{x+1} - \frac{1}{2(x+1)} - \ln(2x+3) + 1\\
&=& \frac{1}{2(x+1)}  -\ln(2x+3).
\end{eqnarray*} 
From here, one finds that the derivate of $\ln\{r_{2}(x+1)/r_{2}(x)\}$ is negative; in consequence, the function $x\mapsto \ln\{r_{2}(x+1)/r_{2}(x)\}$ is decreasing for  $x\in\mathbb{N}$. Besides, given that at $x=0$ we obtain $\ln\{r_{2}(1)/r_{2}(0)\}<0$, it follows that $r_{2}(x+1)/r_{2}(x)< 1$ with $r_{2}(0)=1$.

Comparison of the variance of kernels in equation (\ref{eq:varianceKern}) occurs directly since the discrete standard kernels inherit the intrinsic properties of the discrete distribution from which they were constructed. The binomial distribution is underdispersed (variance $\leq$ mean), the Poisson distribution is equidispersed (variance $=$ mean) and the negative binomial distribution is overdispersed (variance $\geq$ mean). From it comes the ranking of the variance of discrete standard kernels assuming a common mean $\textnormal{E}(\mathcal{K}_{x,h})= x + h$. \hfill$\blacksquare$

\medskip
In the next section, performance of the kernel estimators using discrete standard kernels is investigated according to the properties of their modal probability and variance (highlighted in equations (\ref{eq:probaMod}) and (\ref{eq:varianceKern})). 

\section{Discrete nonparametric kernel estimators} \label{sec3}
This section assesses performance of discrete standard kernel estimators as a global squared error. We rank global squared errors of the estimators studied, which has been previously determined only in numerical simulations \cite{KokonSenga11,Senga08}.

Let $(X_{i})_{i=1,\cdots,n}$ be i.i.d. observations having a p.m.f. $f(\cdot)=\Pr(X_{i}=\cdot)$ to estimate on $\mathbb{N}$. A discrete nonparametric estimator
of $f$ is defined as follows:
\begin{equation}\label{eq:estim_fn}
\widetilde{f}(x)=\frac{1}{n}\sum_{i=1}^{n}K_{x,h}(X_{i})=:\widetilde{f}_{K,h}(x), \text{ \  } x\in\mathbb{N}.
\end{equation}

From \cite{KokonSenga11,Senga08}, the estimator's bias and variance can be decomposed around the target $x\in\mathbb{N}$ such that
\begin{equation*}\label{eq:bias}
\textnormal{Bias}\{\widetilde{f}_{K,h}(x)\}=f(x)\{\Pr(\mathcal{K}_{x,h}=x)-1\} + Q_{n}(x;h)
\end{equation*}
and 
\begin{equation*}\label{eq:var}
\textnormal{Var}\{\widetilde{f}_{K,h}(x)\}=\frac{1}{n}f(x)\{\Pr(\mathcal{K}_{x,h}=x)\}^{2} - \frac{1}{n}f^{2}(x) + R_{n}(x;h),
\end{equation*}
with $$ Q_{n}(x;h)= \sum_{y\in\mathbb{N}\setminus\{x\}}f(y)\Pr(\mathcal{K}_{x,h}=y)$$ and $$R_{n}(x;h)= \frac{1}{n}\sum_{y\in\mathbb{N}\setminus\{x\}}f(y)\{\Pr(\mathcal{K}_{x,h}=y)\}^{2} - \frac{1}{n}\bigg[f(x)+ \sum_{y\in\mathbb{N}}\{f(y)-f(x)\}\Pr(\mathcal{K}_{x,h}=y) \bigg]^{2} + \frac{1}{n}f^{2}(x). $$
The estimator $\widetilde{f}_{K,h}$ is biased since the modal probability of discrete standard kernels does not tend to one when $h$ goes to $0$. A direct consequence of the estimator's bias is the  non-consistency of mean integrated squarred error (MISE) of $\widetilde{f}_{K,h}$ given by
\begin{eqnarray*}\label{eq:mise}
\text{MISE}(\widetilde{f}_{K,h})&=& \sum_{x\in\mathbb{N}}\textnormal{Bias}^{2}\{\widetilde{f}_{K,h}(x)\} + \sum_{x\in\mathbb{N}}\textnormal{Var}\{\widetilde{f}_{K,h}(x)\}\\
&=& \text{AMISE}(\widetilde{f}_{K,h}) + \sum_{x\in\mathbb{N}}[Q_{n}(x;h)+R_{n}(x;h)], 
\end{eqnarray*}
where approximate MISE, called AMISE, corresponds to the leading term such that
\begin{equation}\label{eq:Amise}
\text{AMISE}(\widetilde{f}_{K,h})=\sum_{x\in\mathbb{N}}f^{2}(x)\{\Pr(\mathcal{K}_{x,h}=x)-1\}^{2} + \frac{1}{n}\sum_{x\in\mathbb{N}}f(x)[\{\Pr(\mathcal{K}_{x,h}=x)\}^{2}-f(x)] . \text{ \ \ \ \ }
\end{equation}
For a small/medium sample size, the terms $Q_{n}$ and $R_{n}$ have a non-negligible influence on calculation of $\widetilde{f}_{K,h}$'s bias and variance.  As $n$ increases, $Q_{n}$ becomes smaller but remains different from $0$, and the variance term tends to $0$ since it is penalised by the factor $1/n$. In any case, for discrete standard kernel estimators, we obtain
\[
\sum_{x\in\mathbb{N}}[Q_{n}(x;h)+R_{n}(x;h)] \nrightarrow 0,  \text{ \ as \ } n\to\infty \text{ \ and \ } h\to0. 
\]
However, the decrease in $\widetilde{f}_{K,h}$'s variance term leads to considering mainly the influence of $\widetilde{f}_{K,h}$'s bias term on AMISE.
Note that from equation (\ref{eq:Amise}), the binomial kernel estimator has the lowest approximate integrated squared bias (first term) and the highest approximate integrated variance (second term), while it is the opposite for the negative binomial kernel estimator.  The behaviour of MISE of $\widetilde{f}_{K,h}$ will be illustrated by simulating a known p.m.f. $f$ for several sample sizes $n$ (Section 4.1).

\medskip
\textit{Remark 2}. \textit{(i)} For discrete standard kernels under assumptions ($\text{A}_{3}$)-($\text{A}_{4}$),  equation (\ref{ineq:Amise}) can be found by using an expansion of $\widetilde{f}_{K,h}$'s bias and a majoration of $\widetilde{f}_{K,h}$'s variance as $n\to\infty$ and $h\to0$. By considering the Taylor expansion as $h\to0$, the bias term can be successively expressed as   
\begin{eqnarray}
\textnormal{Bias}\{\widetilde{f}_{K,h}(x)\}&=& \mathbb{E}\{\widetilde{f}(x)\}-f(x)\notag\\
&=&f\{\mathbb{E}(\mathcal{K}_{x,h})\}-f(x) +  \frac{1}{2}\text{Var}(\mathcal{K}_{x,h})f^{(2)}(x)+ o(h), \label{bias_fn} 
\end{eqnarray}
with $f^{(2)}$ being the finite difference of second order of the p.m.f. $f$. Based on the ranking of variance of discrete standard kernels, equation (\ref{bias_fn}) shows that using binomial kernel provides smaller estimator bias than Poisson and negative binomial kernels.  The variance term can be majored as follows:
\begin{equation*}
\textnormal{Var}\{\widetilde{f}_{K,h}(x)\}=\frac{1}{n}\textnormal{Var}\{K_{x,h}(X_{1})\}\leq \frac{1}{n}\mathbb{E}\{K^{2}_{x,h}(x)\},
\end{equation*}
such that we obtain
\[
\textnormal{MISE}(\widetilde{f}_{K,h})=\sum_{x\in\mathbb{N}}\big[f\{\mathbb{E}(\mathcal{K}_{x,h})\}-f(x) +  \frac{1}{2}\text{Var}(\mathcal{K}_{x,h})f^{(2)}(x) \big]^{2} + O\left(\frac{1}{n}\right) +o\left(h^{2}\right),
\]
as $n\to\infty$ large and $h\to0$. Finally, the ranking of MISE of $\widetilde{f}_{K,h}$ results from the ranking of variance of discrete standard kernels, as follows : 
\begin{equation}\label{ineq:mise}
\textnormal{MISE}(\widetilde{f}_{B,h})\leq\textnormal{MISE}(\widetilde{f}_{P,h})\leq\textnormal{MISE}(\widetilde{f}_{NB,h}).
\end{equation}

\textit{(ii)} Since the p.m.f. estimator $\widetilde{f}_{K,h}$ given in equation (\ref{eq:estim_fn}) is not a bona fide estimator ($0<C^{[K]}=\sum_{x\in\mathbb{N}}\widetilde{f}_{K,h}(x)\neq 1$), it required  normalization. The estimator bias has an influence on the behaviour of normalising constant $C$ according to the kernel used such that, for $h\to0$, 
$$ {\color{black}{ \mathbb{E}(C^{[B]}) \leq \mathbb{E}(C^{[P]}) \leq \mathbb{E}(C^{[NB]})}},  $$
since 
\begin{equation*}
\mathbb{E}(C^{[K]})= \sum_{x\in\mathbb{N}}\big[ f(x) + \textnormal{Bias}\{\widetilde{f}_{K,h}(x)\} \big]= 1 + \sum_{x\in\mathbb{N}}\textnormal{Bias}\{\widetilde{f}_{K,h}(x)\}
\end{equation*}
and
\begin{equation*}\label{ineq:bias}
\textnormal{Bias}\{\widetilde{f}_{B,h}(x)\}\leq\textnormal{Bias}\{\widetilde{f}_{P,h}(x)\}\leq\textnormal{Bias}\{\widetilde{f}_{NB,h}(x)\}.
\end{equation*}

\textit{(iii)} Finally, note that the MISE of frequency estimator equals $(1/n)\{1 - \sum_{x\in\mathbb{N}}f^{2}(x) \}$, obtained by assuming that $\Pr(\mathcal{K}_{x,h}=x)=1$ for Dirac type kernel in equation (\ref{eq:Amise}).

Ultimately, we formulate the following proposition:
\begin{proposition}\label{Propostion2}
Consider any fixed $x\in\mathbb{N}$ and $h>0$. As $n\to\infty$ and $h\to0$, the approximate global squared error of the estimators $\widetilde{f}$ using binomial (B), Poisson (P) and negative binomial (NB) kernels satisfy:
\begin{equation}\label{ineq:Amise}
\textnormal{AMISE}(\widetilde{f}_{B,h})\leq\textnormal{AMISE}(\widetilde{f}_{P,h})\leq\textnormal{AMISE}(\widetilde{f}_{NB,h}). 
\end{equation}

\end{proposition}
{\bf Proof}. By using the expression of AMISE in equation (\ref{eq:Amise}), the result is a consequence of the ranking of modal probability of discrete standard kernels in Proposition \ref{Propostion1}. \hfill$\blacksquare$

\section{Illustrations}\label{sec4}
This section illustrates the performance of the nonparametric estimator $\widetilde{f}_{K,h}$ using discrete standard kernels on simulated count data; in addition, applications are proposed for real count data from environmental sciences.

\subsection{Simulations}
We conducted Monte Carlo simulations to compare the discrete kernel estimators using mean values of their bias, variance and global error, but also to investigate effects of sample sizes. Samples were simulated by randomly generating count data from a Poisson p.m.f. $\mathcal{P}(\mu)$ with $\mu=2$. To measure the performance of estimator $\widetilde{f}_{K,h}$ in (\ref{eq:estim_fn}), we used the mean $\text{MISE}$ of $\widetilde{f}_{K,h}$ over $250$ replicates of sample size $n=\{15,25,50,75,100\}$ such that
$$ \overline{\text{MISE}}(\widetilde{f}_{K,h})=\frac{1}{250}\sum_{i=1}^{250}\text{MISE}_{i}(\widetilde{f}_{K,h}),$$ 
with $\text{MISE}_{i}$ being the global squared error of the $\widetilde{f}_{K,h}$ calculated after each replicate $i$ of count data. 

Two main issues of the discrete kernel method are the choices of bandwidth and kernel. Among several procedures, a cross-validation procedure was selected for bandwidth choice{\color{black}{; an example for anoher approach is the Bayesian one \cite{Senga16}}}. Simulations in our study were not time-consuming, and we were essentially interested in ranking the performance of discrete kernel estimators. The cross-validation procedure was satisfying for these aspects, and choosing a different bandwidth-choice procedure did not modify trends in the results.
For each simulation, the smoothing bandwidth was found as $h_{cv}=arg\min_{h>0}CV(h)$ with
\begin{equation*}
CV(h) =\sum_{x\in \mathbb{N}}\left\{ \frac{1}{n}\sum_{i=1}^{n}K_{x,h}\left(
X_{i}\right) \right\} ^{2}-\frac{2}{n(n-1)}\sum_{i=1}^{n}\sum_{j\neq
i}K_{X_{i},h}\left( X_{j}\right)   \notag  \label{CV(h)}
\end{equation*}%
being the cross-validation criterion.\cite{KokonSenga11} For the kernel choice, the non-consistent estimators using discrete standard kernels were compared to the consistent estimator using discrete symmetric triangular kernels $K_{p;x,h}$ (section 2.1). The fixed value $p=1$ was considered, since the MISE of nonparametric estimator $\widetilde{f}_{K,h}$ increases with respect to  $p\in\mathbb{N}$ for a fixed bandwidth $h>0$.\cite{KokonZocchi10} We used the "Ake" package of R software, which uses discrete kernel estimators and a cross-validation procedure \cite{Ake}.

\begin{figure}[!h]
\begin{center}
\includegraphics[height=400pt,width=400pt]{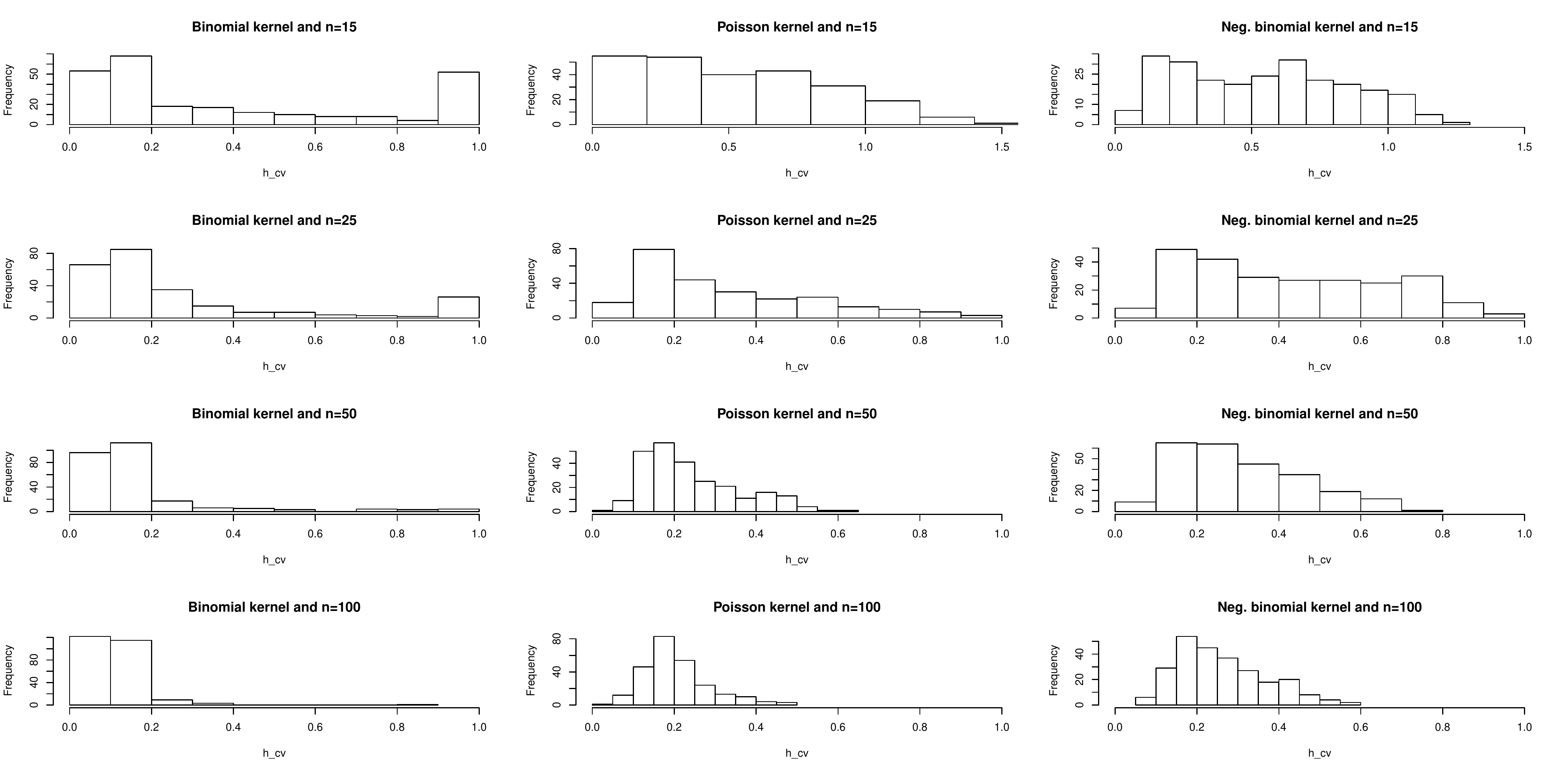}
\end{center}
\caption{Distribution of $h_{cv}$-values for discrete standard kernel estimators of count data of sample sizes $n$ simulated from Poisson p.m.f. $\mathcal{P}(\mu)$ with $\mu=2$.}
\label{fig:hcv}
\end{figure}

{\it Analysis of $h_{cv}$-values.} The distribution of $h_{cv}$-values (Figure \ref{fig:hcv}) and their descriptive statistics (Table \ref{tab:Result_sim_hcv}) confirmed that smoothing parameter values went to $0$ as $n$ increased. For all sample sizes and all discrete standard kernels, the $h_{cv}$-values had an asymmetric distribution with a mean value on the left (closer to $0$) and the tail of the distribution on the right. Due to having a smoothing parameter defined on the interval (0,1], the binomial kernel estimator had mean $h_{cv}$ values smaller than those of other discrete kernel estimators, including those of the discrete symmetric triangular kernel (Table \ref{tab:Result_sim_hcv}).

\begin{table}[htbp]
\caption{Mean and standard deviation (sd) of $h_{cv}$-values for discrete kernel
estimators of count data simulated from Poisson p.m.f. $\mathcal{P}(\mu)$ with $\mu=2$. }
\scriptsize
\center
{\begin{tabular}[l]{@{}crrrrrrrr}
\multicolumn{9}{c}{} \\ \hline
Sample  &  \multicolumn{2}{c}{Neg. bin.  kern.} &  \multicolumn{2}{c}{Pois. kern.}  &  \multicolumn{2}{c}{Bin. kern.}  &  \multicolumn{2}{c}{Triang. $a=1$ kern.}  \\ 
size $n$ & \multicolumn{2}{c}{estimator} &\multicolumn{2}{c}{estimator} & \multicolumn{2}{c}{estimator}& \multicolumn{2}{c}{estimator} \\ \hline 
& mean & sd & mean & sd& mean& sd & mean & sd\\
15&  $0.55$ & $0.301$ & $0.53$ & $0.345$& $0.40$ & $0.360$ & $1.75$ &    $0.962$\\
25&   $0.43$& $0.232$ &$0.33$ & $0.217$& $0.28$& $0.287$&  $1.89$&   $1.074$\\
50&   $0.31$& $0.149$& $0.25$& $0.117$& $0.17$& $0.175$& $1.87$&  $1.193$\\
75&   $0.26$ & $0.109$ & $0.21$& $0.080$& $0.11$ & $0.067$& $1.81$&   $1.264$\\
100&   $0.23$& $0.086$ & $0.18$& $0.051$ & $0.09$& $0.032$& $1.62$&  $1.268$ \\ \hline
\end{tabular}}
\label{tab:Result_sim_hcv}
\end{table}

{\it Peformance of the estimators in terms of bias and variance.} Table \ref{tab:Result_sim} presents mean integrated squared bias ($\overline{\text{IBias}}$), integrated variance ($\overline{\text{IVar}}$) and MISE ($\overline{\text{MISE}}$). On average, the binomial kernel estimator had lower integrated squared bias than the two other discrete standard kernel estimators but higher integrated variance, while it was the opposite for the negative binomial kernel estimator. Thus, the binomial kernel estimator outperformed the Poisson and negative binomial estimators in term of bias, while the negative binomial estimator was the most effective in terms of variance (or standard deviation).  Only the discrete triangular kernel estimator had low values of both $\overline{\text{IBias}}$ (close to those of the binomial kernel estimator) and $\overline{\text{IVar}}$  (close to those of the negative binomial kernel estimator). 

{\it  Effect of sample sizes.}  For sample sizes $n=\{15, 25\}$, comparison of resulting $\overline{\text{MISE}}$ of discrete standard kernel estimators  showed that the Poisson kernel estimator outperformed  the binomial and negative binomial kernel estimators of the simulated p.m.f. Also, for the smallest sample size considered $n=15$, the negative binomial kernel estimator was even better than binomial kernel estimator. For $n\geq 50$, the binomial kernel estimator became better than the two other discrete standard kernel estimators. The sample size $n=50$ corresponded to the maximum $n_{0}$, described in subsection \ref{ssec2:comparaison_kernel}, which defined the domain of relative efficiency of the kernels. Finally, for all sample sizes considered, the discrete triangular kernel estimator provided the best fit to the simulated count data. For sample sizes $n=\{75,100\}$, however, the binomial and discrete triangular kernel estimators had similar performances. Compared to the frequency estimator $\widetilde{F}$, all discrete kernel estimators considered provided smaller MISE than $\widetilde{F}$ for $n=\{25,50\}$, and only binomial and discrete triangular kernel estimators provided smaller MISE than $\widetilde{F}$ for all sample sizes. 

\begin{table}[ht]
\caption{Results of average mean integrated squared error ($\overline{\text{MISE}}$), integrated squared bias ($\overline{\text{IBias}}$) and integrated variance ($\overline{\text{IVar}}$) for discrete kernel
estimators of count data simulated from Poisson distribution with mean 2. Results are multiplied by $10^3$.}
\scriptsize
\center
{\begin{tabular}[l]{@{}crrrrrrrrrrrrr}\hline
Sample  & Dirac &  \multicolumn{3}{c}{Neg. bin. kern.} &  \multicolumn{3}{c}{Pois. kern.} &  \multicolumn{3}{c}{Bin. kern.} &  \multicolumn{3}{c}{Triang. $p=1$ kern.} \\ 
size $n$& kern. &\multicolumn{3}{c}{estimator}  & \multicolumn{3}{c}{estimator} & \multicolumn{3}{c}{estimator} & \multicolumn{3}{c}{estimator}\\\hline 
& $\overline{\text{MISE}}$& $\overline{\text{IBias}}$ & $\overline{\text{IVar}}$& $\overline{\text{MISE}}$& $\overline{\text{IBias}}$& $\overline{\text{IVar}}$&  $\overline{\text{MISE}}$ & $\overline{\text{IBias}}$ & $\overline{\text{IVar}}$& $\overline{\text{MISE}}$& $\overline{\text{IBias}}$& $\overline{\text{IVar}}$&  $\overline{\text{MISE}}$\\
15& $52.8$ &$26.8$ & $4.5$ &$30.9$ &$18.5$ &$4.8$ & $24.0$& $14.3$ & $18.3$ & $32.7$ & $3.1$ & $11.3$ & $15.4$\\
25& $31.7$ & $24.5$ & $3.1$ &$27.5$ & $14.7$ & $3.6$ & $18.0$ & $9.4$ & $9.8$& $18.9$ & $2.4$& $7.7$ & $9.7$\\
50& $15.8$& $24.0$& $1.7$ & $25.8$ & $13.2$&$2.0$ & $15.2$ &  $4.0$ & $4.3$ & $7.9$ & $2.0$ & $3.9$&$6.2$\\
75& $10.6$& $24.2$& $1.2$ & $25.5$& $13.0$& $1.4$& $14.4$& $2.5$ & $2.7$ & $5.3$ & $1.9$ & $2.8$& $4.8$\\
100& $7.9$& $24.1$& $0.9$& $25.2$& $12.9$ &$1.4$ &$14.1$ & $2.4$ & $2.1$ & $4.5$ & $1.8$ & $2.2$ &$4.1$\\ \hline
\end{tabular}}
\label{tab:Result_sim}
\end{table}

\subsection{Applications}
The real data sets were explanatory count variables describing development of an insect pest (spiralling whitefly, \textit{Aleurodicus dispersus} Russel), which damages plants by sucking sap, decreasing photosynthesic activity and drying up leaves. This insect, originally from Central America and the Caribbean, is present in Congo-Brazzaville, and Congolese biologists were seeking to model its development.  Thus,  experimental plantations were established for several host plants, such as the fruit trees known as safou (\textit{Dacryodes edulis}) and huru (\textit{Hura crepitans}). Among other data collected, pre-adult developement time was quantified as the number of days required for an insect to develop from egg to adult stages (Table \ref{tab:data}). The medium sample size $n=\{51, 60\}$ was one reason for choosing these data sets to illustrate the utility of non-consistent discrete kernel estimators. 

\begin{table}[ht]
\caption{Observed pre-adult development time (days) of spiraling whitefly observed on two species of fruit trees}
\scriptsize
\center
{\begin{tabular}[l]{@{}cccccccccccccc}\hline
\multicolumn{13}{c}{\textsc{Safou tree} } & Total $n$\\ \hline
Development time (days)& $30$ & $31$ & $32$ & & & & & & & & & &\\
Number of insects observed &$28$ &$21$ &$11$ & & & & & & & & & & $60$\\ \hline

\multicolumn{13}{c}{\textsc{Hura tree} } &\\ \hline
Development time (days)& $25$ &$26$ & $27$ & $28$ &$29$ &$30$ &$31$ & $32$& $33$&$34$ &$35$ & &\\
Number of insects observed & $5$& $5$& $7$ & $8$ & $11$ & $2$ & $1$ &$4$ &$4$ &$2$ &$2$ & & $51$\\ 
\hline
\end{tabular}}
\label{tab:data}
\end{table}

Nonparametric estimators using discrete standard kernels and discrete symmetric triangular kernel with $p=1$ were applied to count data (Table \ref{tab:data}). The bandwidth parameter was selected using the cross-validation procedure. Performance of nonparametric discrete kernel estimators $\widetilde{f}_{K,h}$ of empirical frequency $f_{0}$ of the count data studied was assessed using the practical integrated squared error (ISE), given as
\begin{equation*}
\textnormal{ISE}(h)=\sum_{x\in \mathbb{N}}\left\{ \widetilde{f}_{n,{K}_{x,h}}(x)-f_{0}(x)\right\}.
^{2}
\end{equation*}%
Note that in this case there were few alternatives to using the ISE criterion based on a Dirac kernel estimator ($f_{0}$), which is a poor estimator on its own. 

\begin{figure}[!h]
\begin{center}
\includegraphics[width=\textwidth]{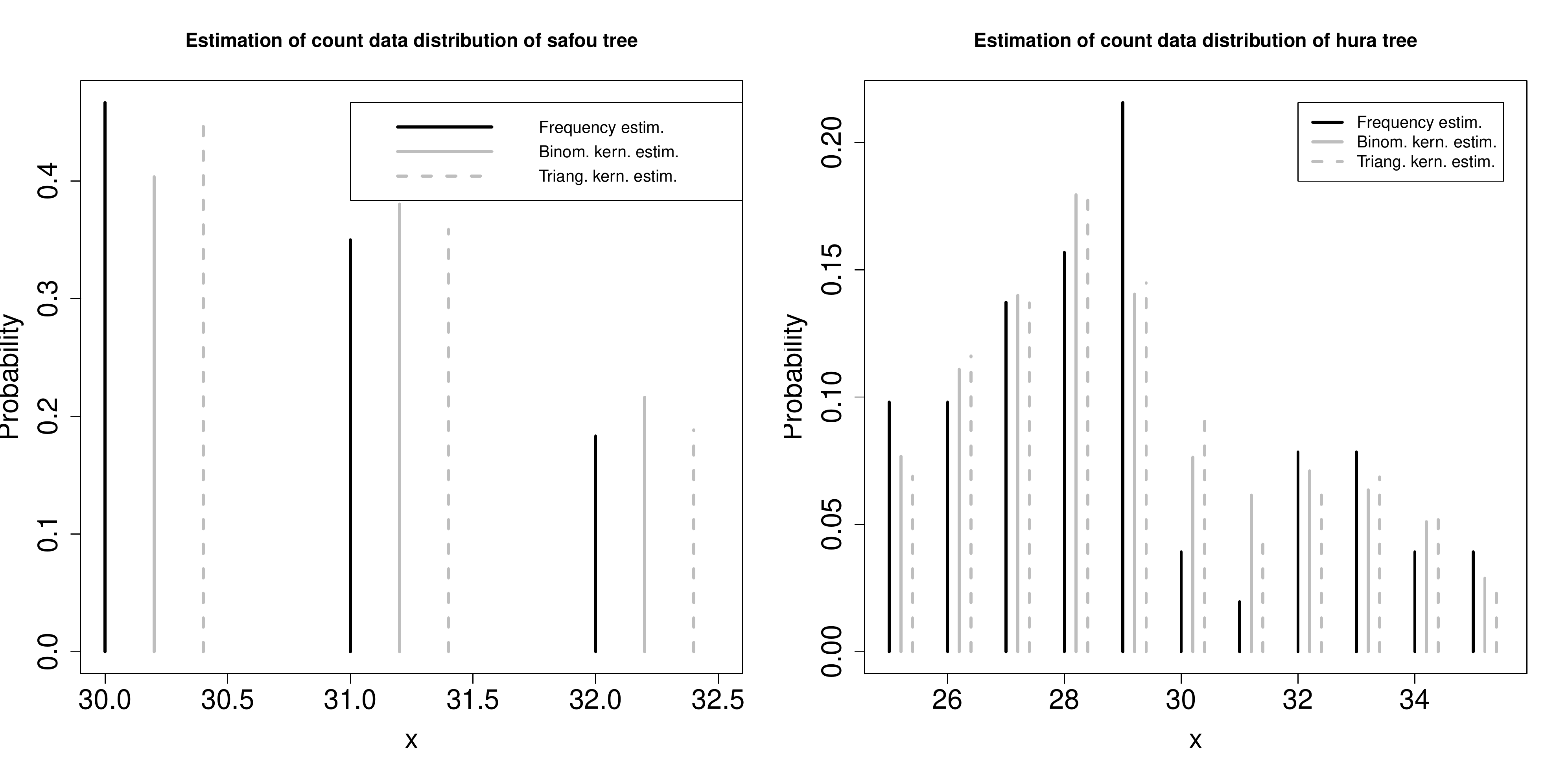}
\end{center}
\caption{Nonparametric kernel estimates of empirical frequency of count data from Table \ref{tab:data} }
\label{fig:estimation_data_tree}
\end{figure}

{\it Peformance of the estimators in terms of ISE.} The discrete symmetric triangular kernel estimator performed better than discrete standard kernel estimators for adjusted count data of insects on the safou tree, while the binomial kernel estimator performed better than all other discrete kernel estimators studied for adjusted count data of insects on the hura tree (Table \ref{tab:result_data}). In two cases, the lowest-performing estimators were Poisson and negative binomial kernel estimators. Figure \ref{fig:estimation_data_tree} presents discrete binomial and symmetric triangular kernel estimates.

\begin{table}[ht]
\caption{ISE and $h_{cv}$-values from nonparametric kernel estimates of empirical frequency of data in Table \ref{tab:data}. Bold values indicate the smallest ISE. }
\scriptsize
\center
{\begin{tabular}[l]{@{}ccccc}\hline
&Neg. bin kern. & Pois. kern. & Bin. kern. & Triang. $p=1$ kern.\\
& estimator & estimator & estimator & estimator \\ \hline
Safou tree & $0.0408$ ($h_{cv}=0.05$)&  $0.0382$ ($h_{cv}=0.08$) & $0.0059$ ($h_{cv}=0.004$) & ${\bf 0.0003}$ ($h_{cv}=0.08$)\\ 
Hura tree & $0.0305$ ($h_{cv}=0.75$) & $0.0261$ ($h_{cv}=0.87$) & ${\bf 0.0104}$ ($h_{cv}=0.02$)& $0.0112$ ($h_{cv}=4.65$)\\
\hline
\end{tabular}}
\label{tab:result_data}
\end{table}

Concluding these application cases, count data distribution was displayed for which the non-consistent binomial kernel estimator may be more appropriate than the consistent discrete symmetric triangular kernel estimator. The small difference between the binomial and discrete triangular kernel estimators in these cases suggest that either can be applied for smoothing count data of medium sample size.

\section{Concluding remarks}
This work seeks to contribute to better understanding of the discrete associated kernel method for estimating count data. The main difference emphasised between the discrete kernels comes  from the behaviour of both their modal probability and variance. Ranking the performance of nonparametric estimators using discrete standard asymmetric kernels showed that the binomial kernel estimator generally outperformed the two other discrete kernel estimators for medium or larger sample sizes, in terms of global squared error. The simulation study confirmed the previous ranking and also showed that the consistent discrete symmetric triangular kernel estimator generally outperforms the non-consistent discrete standard asymmetric kernel estimators. Nevertheless, the application case displayed a count data distribution with medium sample size in which the binomial kernel estimator may be better or equivalent to the discrete triangular kernel estimator. The question remains of the maximum value of sample size and/or the smoothing parameter to define the domain of the relative efficiencies of discrete standard kernels. Finally, discrete nonparametric kernel estimation is confirmed to be a valuable alternative to empirical estimation of count data distribution, specially for small/medium sample sizes, as previously noted by \cite{KokonSenga11,Senga08}. An interesting perspective would be to establish a performance criterion to compare the relative efficiency of any discrete kernel to that of the Dirac type kernel. 

\section*{Acknowledgments}
The author thanks two anonymous referees and the Associate Editor for their careful review and
helpful comments that led to considerable improvement of the article. The author is also grateful to Michael Corson for helping improve the English content.


\begin{center}
{\bf BIBLIOGRAPHY}
\end{center}
\begin{enumerate}
\bibitem{KokonSenga11} Kokonendji CC, Senga Kiess\'e T. Discrete associated kernel method and extensions. Statistical Methodology. 2011;8:497-516.
\bibitem{Senga08} Senga Kiess\'e T. Nonparametric approach by discrete associated kernel for count data. Ph.D. of University of Pau France; 2008.
\bibitem{Simonoff96} Simonoff JS. Smoothing methods in statistics. Springer New York; 1996.
\bibitem{Tsyb04} Tsybakov AB. Introduction \`a l'estimation non-param\'etrique. Springer Paris; 2004.
\bibitem{AitchAitk76} Aitchison J, Aitken CGG. Multivariate binary discrimination by the kernel method. Biometrika. 1976;63:413-420.
\bibitem{Senga16} Senga Kiess\'e T, Zougab N, Kokonendji CC. Bayesian estimation of bandwidth in semiparametric
kernel estimation of unknown probability mass and regression functions of count data. Computational Statistics. 2016;31:189-206.
\bibitem{Zougab13} Zougab N, Adjabi S, Kokonendji CC. Adaptive smoothing in associated kernel discrete functions estimation using bayesian approach. Journal of Statistical Computation and Simulation. 2013;83:2219-2231.
\bibitem{Belaid16} Belaid N, Adjabi S, Zougab N, Kokonendji CC. Bayesian bandwidth selection in discrete multivariate associated kernel estimators for probability mass functions. Journal of the Korean Statistical Society. 2016;45:557-567.
\bibitem{KokonZocchi10} Kokonendji CC, Zocchi SS. Extensions of discrete triangular distribution and boundary bias in kernel estimation for discrete functions. Statistical and Probability Letters. 2010; 80:1655-1662.
\bibitem{WangVan81} Wang MC, Ryzin JV. A class of smooth estimators for discrete distributions. Biometrika. 1981;68:301-309.
\bibitem{Senga14} Senga Kiess\'e T, Lorino T, Khraibani H. Discrete nonparametric kernel and parametric methods for the modeling of pavement deterioration. Communications in Statistics - Theory and Methods. 2014;43:1164-1178.
\bibitem{Chen99} Chen SX. A beta kernel estimator for density functions. Computational Statistics and Data Analysis. 2000;31:131-145.
\bibitem{HagmScail07} Hagmann M, Scaillet M. Local multiplicative bias correction for asymmetric kernel density estimators. Journal of Econometrics. 2007;141:213-249.
\bibitem{Ake} Wansouw\'e WE, Som\'e SM, Kokonendji CC. Ake: Associated kernel estimations. 2015; r package version 1.0; Available from: http://CRAN.R-project.org/package=Ake.
\end{enumerate}
\end{document}